\documentclass[twocolumn,10pt]{IEEEtran}
\usepackage{ifpdf, flushend}

\ifCLASSINFOpdf
  \usepackage[pdftex]{graphicx}
  \graphicspath{{../pdf/}{../jpeg/}}
  \DeclareGraphicsExtensions{.pdf,.jpeg,.png}
\else
  \usepackage[dvips]{graphicx}
  \graphicspath{{../eps/}}
  \DeclareGraphicsExtensions{.eps}
\fi
\usepackage[cmex10]{amsmath}
\usepackage {amssymb}
\usepackage{algorithmic}
\usepackage{array}
\usepackage{mdwmath}
\usepackage{mdwtab}
\usepackage{eqparbox}
\usepackage{url}
\usepackage{hyperref}
\usepackage{algorithm}
\usepackage{algorithmic}
\usepackage{epstopdf,cite,color}

\usepackage{amsthm} 
\newtheorem{theorem}{Theorem}
\newtheorem{lemma}{Lemma}

\newcommand{\beq}{\begin{equation}}
\newcommand{\eeq}{\end{equation}}
\newcommand{\beqn}{\begin{eqnarray}}
\newcommand{\eeqn}{\end{eqnarray}}
\newcommand{\beqno}{\begin{eqnarray*}}
\newcommand{\eeqno}{\end{eqnarray*}}
\newcommand{\bma}{\begin{displaymath}}
\newcommand{\ema}{\end{displaymath}}
\newcommand{\bnu}{\begin{enumerate}}
\newcommand{\enu}{\end{enumerate}}
\newcommand{\bce}{\begin{center}}
\newcommand{\ece}{\end{center}}
\newcommand{\btb}{\begin{tabular}}
\newcommand{\etb}{\end{tabular}}

\begin{document}

%
\title{Power Control and Relay Selection in Full-Duplex Cognitive Relay Networks: Coherent versus Non-coherent Scenarios}

\author{Le~Thanh~Tan,~\IEEEmembership{Member,~IEEE,} Lei~Ying,~\IEEEmembership{Member,~IEEE,} and Daniel W. Bliss,~\IEEEmembership{Fellow,~IEEE}
\thanks{The authors are with the School of Electrical, Computer and Energy Engineering, Arizona State University,  Tempe, AZ 85287, USA.
Emails: \{tlethanh,lei.ying.2,d.w.bliss\}@asu.edu.}}

\maketitle
\thispagestyle{empty}

\begin{abstract}
This paper investigates power control and relay selection in Full Duplex Cognitive Relay Networks (FDCRNs), where the secondary-user (SU) relays can simultaneously receive and forward the signal from the SU source. We study both non-coherent and coherent scenarios. In the non-coherent case, the SU relay forwards the signal from the SU source without regulating the phase; while in the coherent scenario, the SU relay regulates the phase when forwarding the signal to minimize the interference at the primary-user (PU) receiver. We consider the problem of maximizing the transmission rate from the SU source to the SU destination subject to the interference constraint at the PU receiver and power constraints at both the SU source and SU relay. We develop low-complexity and high-performance joint power control and relay selection algorithms. The superior performance of the proposed algorithms are confirmed using extensive numerical evaluation. In particular, we demonstrate the significant gain of phase regulation at the SU relay (i.e., the gain of the coherent mechanism over the noncoherent mechanism).

\end{abstract}

\IEEEpeerreviewmaketitle

\section{Introduction}

Cognitive radio is one of the most promising technologies for addressing today's spectrum shortage \cite{Tan11,Liang11}. This paper considers {\em underlay} cognitive radio networks where primary and secondary networks transmit simultaneously over the same spectrum under the constraint that the interference caused by the secondary network to the primary network is below a pre-specified threshold \cite{Le08}. 
In particular, we consider a cognitive relay network where the use of SU relay can significantly increase the transmission rate because of path loss reduction. 
Most existing research on underlay CRNs has focused on the design and analysis of cognitive relay networks with half-duplex (HD) relays  \cite{Liang11}.

Different from these existing work, this paper considers full-duplex relays, which can transmit and receive simultaneously on the same frequency band \cite{Day12b, Bliss07}. Comparing with HD relays, FD relays can achieve both higher throughput and lower latency with the same amount of spectrum. 
Design and analysis of FDCRNs, however, are very different from HDCRNs due to the presence of {\em self-interference}, resulted from the power leakage from the transmitter to the receiver of a FD transceiver. 
The FD technology can improve  spectrum access efficiency in cognitive radio networks \cite{Tan15a,Tan15, Tan16} where SUs can sense and transmit simultaneously. 
However these results assume the interweave spectrum sharing paradigm under which SUs only transmit when PUs are not transmitting.
Moreover, engineering of a cognitive FD relaying network has been considered in \cite{Kim12, Kim15}, where various resource allocation algorithms to improve the outage probability have been proposed.
These existing results focus on either minimizing the outage probability or analyzing performance for existing algorithms.

This paper focuses on power control and relay selection in FDCRNs with explicit consideration of self-interference. 
We assume SU relays use the amplify-and-forward (AF) protocol, and further assume full channel state information in both the non-coherent and coherent scenarios and the transmit phase information in the coherent scenario.
We first consider the power control problem in the non-coherent scenario. We formulate the rate maximization problem where the objective is the transmission rate from the SU source to the SU destination, and the constraints include the power constraints at the SU source and SU relay and the interference constraint at the PU receiver.
The rate maximization problem is a non-convex optimization problem. However, it becomes convex if we fix one of two optimization variables.  
Therefore, we propose an alternative optimization algorithm to solve the power control problem.
After calculating the achievable rate for each FD relay, the algorithm selects the one with the maximum rate.

We then consider the coherent scenario, where in addition to control the transmit power, a SU relay further regulates the phase of the transmitted signal to minimize the interference at the PU receiver. We also formulate a rate maximization problem, which again is nonconvex. For this coherent scenario, we first calculate the phase to minimize the interference at the PU receiver. Then we prove that the power-control problem becomes convex when we fix either the transmit power of the SU source then optimize the transmit power of the SU relay or vice versa. We then propose an alternative optimization method for power control. 
Extensive numerical results are presented to investigate the impacts of different parameters on the SU network rate performance and the performance of the proposed power control and relay selection algorithms. From the numerical study, we observe significant rate improvement of FDCRNs compared with HDCRNs. Furthermore, the coherent mechanism yields significantly higher throughput than that under the non-coherent mechanism.

\section{System Model}
\label{SystemModel}

\begin{figure}[!t]
\centering
\includegraphics[width=85mm]{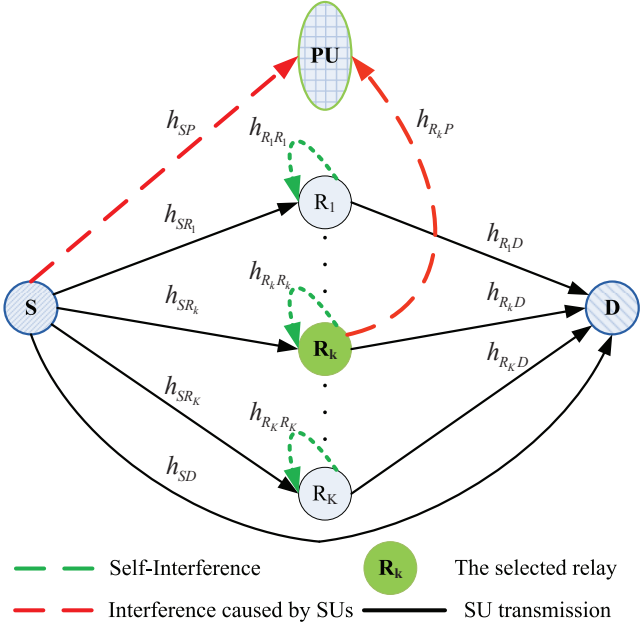}
\caption{System model of power allocation with relay selection for the cognitive full-duplex relay network.}
\label{SCRSN}
\end{figure}

We consider a cognitive relay network which consists of one SU source $S$, $K$ SU relays $R_k$ ($k = 1, \ldots, K$), one SU destination $D$, and one PU receiver $P$ as shown in Fig.~\ref{SCRSN}.
The SU relays are equipped with FD transceivers to work in the FD mode.
Therefore the receiver performance of each SU relay is affected by the self-interference from its transmitter since the transmit power is leaked into the received signal.

Each SU relay $R_k$ uses the AF protocol, and amplifies the received signal from $S$ with a variable gain $G_k$ and forwards the resulting signal to SU destination, $D$.
We denote $h_{SR_k}$, $h_{R_kD}$, $h_{SD}$, $h_{R_kP}$ and $h_{R_kR_k}$ by the corresponding channel coefficients of links $S \rightarrow R_k$, $R_k \rightarrow D$, $S \rightarrow D$, $R_k \rightarrow P$ and $R_k \rightarrow R_k$.
Let  $P_S$ denote the transmit power of SU source $S$.
We also denote by $x_S(t)$, $y_{R_k}(t)$ and $y_D(t)$ the generated signal by the SU source, the transmitted signals at the SU relay and the received signals at the SU destination, respectively.

\begin{figure}[!t] 
\centering
\includegraphics[width=90mm]{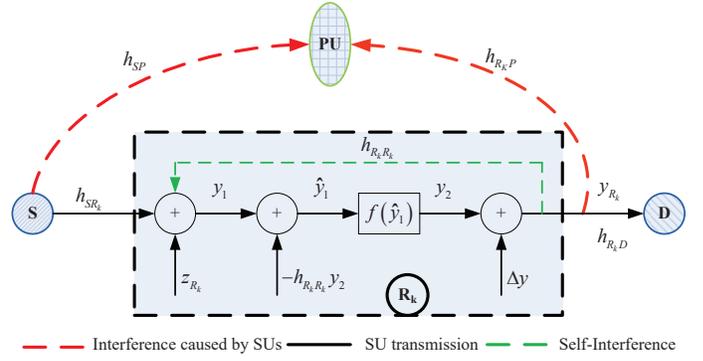}
\caption{The process at FD relay $R_k$.}
\label{SRkCRSN}
\end{figure}

Let us consider a specific SU relay (say relay $R_k$). Fig.~\ref{SRkCRSN} illustrates the signal processing at the relay.
At time $t$, the received signals at SU relay $R_k$ and SU destination $D$ are 
\beqn
y_1(t) \!=\! h_{SR_k} \sqrt{P_S} x_S(t) \!+\! h_{R_kR_k} \left(y_2(t) \!+ \! \Delta y(t)\right) \!+\! z_{R_k}(t) \label{EQN_receiveRD1}\\
y_D(t) \!=\! h_{R_kD} y_{R_k}(t) + h_{SD} \sqrt{P_S} x_S(t) + z_D (t), \label{EQN_receiveRD2} \hspace{1.2cm}
\eeqn
where $z_{R_k}(t)$ and $z_D (t)$ are the additive white Gaussian noises (AWGN) with zero mean and variances $\sigma_{R_k}^2$ and  $\sigma_D^2$, respectively;
$y_D(t)$ and $y_1(t)$ are the received signals at SU relay $R_k$ and SU destination $D$; and
$y_2(t)$ is the received signal after the amplification.
In the following, we ignore the direct signal from the SU source to the SU destination (i.e., the second part in equation (\ref{EQN_receiveRD2})). Note that this assumption is has been used in the literature \cite{Day12b, Kim15} when there is attenuation on the direct transmission channel.

The transmitted signals at SU relay $R_k$ is $$y_{R_k}(t) = y_2(t) + \Delta y(t),$$ where $y_2(t) = f\left(\hat{y}_1\right) = G_k \hat{y}_1 (t-\Delta).$
We should note that the SU relay amplifies the signal by a factor of $G_k$ and delays with duration of $\Delta$.
In the noncoherent scenario, $\Delta$ is fixed. In the coherence scenario, the delay $\Delta$ will be optimized  to minimize the interference at the PU receiver. Furthermore, $\Delta y(t)$ is the noise and follows the i.i.d. Gaussian distribution with zero mean and variance of $P_{\Delta} = \zeta P_{R_k}$ \cite{Day12b, Bliss07}.
$G_k$ can be expressed as $$G_k = \left[P_S \left|h_{SR_k}\right|^2 + \zeta P_{R_k} \left|h_{R_kR_k}\right|^2 + \sigma_{R_k}^2\right]^{-1/2}.$$
We assume that the channel $h_{R_kR_k}$ is perfectly estimated and hence the received signal after self-interference cancellation is
\beqn
\hat{y}_1(t) \!= \sqrt{P_{R_k}} \left(y_1(t) - h_{R_kR_k} y_2(t)\right) \hspace{2.7cm} \nonumber\\
= \sqrt{P_{R_k}} \! \left[\! h_{SR_k} \! \sqrt{P_S} x_S(t) \!+\! h_{R_kR_k} \Delta y(t) \!+\! z_{R_k}(t) \!\right].
\eeqn
In the equation above, $y_2(t)$ is known at SU relay $R_k$ and therefore is used to cancel the interference.
However, the remaining $h_{R_kR_k} \Delta y(t)$ is still present at the received signals and is called the residual interference.
So we can write the transmitted signals at SU relay $R_k$ as follows:
\beqn
y_{R_k}(t) = G_k h_{SR_k} \sqrt{P_{R_k}} \sqrt{P_S} x_S(t-\Delta) + \Delta y(t) \hspace{1cm} \nonumber\\
\!+ G_k h_{R_kR_k} \!\sqrt{P_{R_k}} \Delta y(t\!-\!\Delta) \!+\! G_k \!\sqrt{P_{R_k}} z_{R_k}(t\!-\!\Delta). \label{EQN_y_Rk_trans}
\eeqn

\section{Power Control and Relay Selection}
\label{PAFRS_Problem}

In this section, we study the problem of maximizing the rate between SU source and SU destination while protecting the PU via power control and relay selection.

\subsection{Problem Formulation}
\label{RateOpt}

Let $\mathcal{C}_k(P_S, P_{R_k})$ denote the achieved rate of the FDCRN with relay $R_k,$ which is the function of transmit power of SU source $S$ and transmit power of SU relay $R_k$.
Assume the interference caused by the SU source and relay, $\mathcal{I}_k$ is required to be at most $\overline{\mathcal{I}}_{P}$ to protect the PU.

Now, the rate maximization problem for the selected relay $R_k$ can be stated as follows:

\noindent
\textbf{Problem 1:}
\begin{equation}
\label{EQN_OPTRS}
\begin{array}{l}
 {\mathop {\max }\limits_{P_S, P_{R_k}}} \quad \mathcal{C}_k(P_S, P_{R_k})  \\
 \mbox{s.t.}\,\,\,\, \mathcal{I}_k\left(P_S, P_{R_k}\right) \leq \mathcal{\overline I}_{P}, 0 \leq P_S \leq P_S^{\sf max},\\
 \quad \quad 0 \leq P_{R_k} \leq P_{R_k}^{\sf max},\\
 \end{array}\!\!
\end{equation}
where $P_S^{\sf max}$ and $P_{R_k}^{\sf max}$ are the maximum power levels for the SU source and SU relay, respectively.
The first constraint on $\mathcal{I}_k\left(P_S, P_{R_k}\right)$ requires that the interference caused by the SU transmission is limited.
Moreover, the SU relay's transmit power $P_{R_k}$ must be appropriately set to achieve good tradeoff between the rate of the SU network and self-interference mitigation.

Then the relay selection is determined by
\beqn
\label{EQN_OPT_RELAY_SELEC}
k^* = {\mathop  {\arg \max }\limits_{k \in \left\{1,\ldots, K\right\}}} \quad \mathcal{C}_k^*
\eeqn
where $\mathcal{C}_k^*$ is the solution of (\ref{EQN_OPTRS}).
In the following, we show how to calculate the achieved rate, $\mathcal{C}_k(P_S, P_{R_k})$ and the interference imposed by SU transmissions, $\mathcal{I}_k\left(P_S, P_{R_k}\right)$.

\subsection{The Achievable Rate}
\label{Rate_Formu}

When SU relay $R_k$  is selected, the achievable rate of the link $S \rightarrow R_k \rightarrow D$ based on (\ref{EQN_receiveRD1}) and (\ref{EQN_receiveRD2}) is as follows:
\beqn
\mathcal{C}_k = \log_2 \left[1+ \frac{\frac{P_{R_k} \left|h_{R_kD}\right|^2}{\sigma_D^2} \frac{P_S \left|h_{SR_k}\right|^2}{\hat{\zeta} P_{R_k}+\sigma_{R_k}^2}}  {\mathcal{A}}\right]
\eeqn
where
\beqn
\mathcal{A} &=& 1+ \frac{P_{R_k} \left|h_{R_kD}\right|^2}{\sigma_D^2} + \frac{P_S \left|h_{SR_k}\right|^2}{\hat{\zeta} P_{R_k}+\sigma_{R_k}^2} \label{EQN_CAL_A}\\
\hat{\zeta} &=& \left|h_{R_kR_k}\right|^2 \zeta \label{EQN_zeta_hat}
\eeqn
Recall that we assume the direct signal from the SU source to the SU destination is negligible.

\subsection{The Imposed Interference at PU}
\label{Interference_Formu}

We now determine the interference at the PU caused by the CRN. The interference is the signals from the SU source $S$ and the selected relay $R_k:$
\beqn
\label{EQN_Inter}
y_I^{\sf PU} (t) \!=\! h_{SP} \sqrt{P_S} x_S(t) \!+\! h_{R_kP} y_{R_k}(t)\! +\! z_P(t)
\eeqn
where $z_P(t)$ is the AWGN with zero mean and variance $\sigma_P^2,$ and $y_{R_k}(t)$ is defined in (\ref{EQN_y_Rk_trans}).

We next derive and analyze the interference in two cases: coherent and non-coherent. In particular, we focus on coherent/non-coherent transmissions from the SU source and the SU relay to the PU receiver. All other transmissions are assumed to be non-coherent.  In the coherent scenario, the phase information is needed in the coherent mechanism. This information can be obtained by using methods such as the implicit feedback (using reciprocity between forward and reverse channels in a time-division-duplex system), and explicit feedback (using feedback in a frequency-division-duplex system) \cite{Mudumbai09} or the channel estimation \cite{Arslan07}.

\subsubsection{Non-coherent Scenario}

From (\ref{EQN_Inter}) and (\ref{EQN_y_Rk_trans}), the received interference at the PU caused by the SU source and the selected relay can be written as follows:
\beqn
\mathcal{I}^{\sf non}_k\left(P_S, P_{R_k}\right) = \left|h_{SP}\right|^2 P_S + \left|h_{R_kP}\right|^2 \zeta P_{R_k} \hspace{1.5cm}\nonumber\\
+ G_k^2 \left|h_{R_kP}\right|^2 \! P_{R_k} \left[ \left|h_{SR_k}\right|^2 \! P_S \!+\! \left|h_{R_kR_k}\right|^2 \! \zeta P_{R_k} \!+\! \sigma_{R_k}^2\right]
\eeqn
After simple calculation, we obtain
\beqn
\mathcal{I}^{\sf non}_k\left(P_S, P_{R_k}\right) = \left|h_{SP}\right|^2 P_S + \left|h_{R_kP}\right|^2 P_{R_k} \left(1+\zeta\right)
\eeqn

\subsubsection{Coherent Scenario}

Combining (\ref{EQN_Inter}) with (\ref{EQN_y_Rk_trans}),  the received interference at the PU is
\beqn
\label{EQN_I_k_coh_ori}
\mathcal{\bar{I}}^{\sf coh}_k\left(P_S, P_{R_k}, \phi\right) = \left|A+B e^{-j\phi}\right|^2
\eeqn
where
\beqn
A = h_{SP} \sqrt{P_S} + h_{R_kP}\sqrt{\zeta P_{R_k}} = \left|A\right| \angle{\phi_A}
\eeqn
\beqn
B = \left(h_{SR_k} \sqrt{P_S} + h_{R_kR_k} \sqrt{\zeta P_{R_k}} + \frac{\sigma_{R_k}}{\sqrt{2}} (1+j)\right) \nonumber\\
\times G_k h_{R_kP} \sqrt{P_{R_k}} = \left|B\right| \angle{\phi_B}
\eeqn
and $\phi = 2 \pi f_s \Delta$, $f_s$ is the sampling frequency.

Before using $\mathcal{\bar{I}}^{\sf coh}_k\left(P_S, P_{R_k}, \phi\right)$ in the constraint of the optimization problem, we can minimize $\mathcal{\bar{I}}^{\sf coh}_k\left(P_S, P_{R_k}, \phi\right)$ over the variable $\phi$ at given $\left(P_S, P_{R_k}\right)$, i.e.,
\noindent
\begin{equation}
\label{EQN_OPT_PHI}
 {\mathop {\min }\limits_{\phi} \quad \mathcal{\bar{I}}^{\sf coh}_k\left(P_S, P_{R_k}, \phi\right) }.
\end{equation}

\vspace{0.2cm}
\noindent

\begin{theorem} \label{theorem1}
The optimal solution to (\ref{EQN_OPT_PHI}) is
\beqn
\label{EQN_I_COH}
\phi_{\sf opt} = \pi + \phi_B - \phi_A \hspace{3cm} \nonumber \\
\mathcal{I}^{\sf coh}_k\left(P_S, P_{R_k}\right) =  \mathcal{\bar{I}}^{\sf coh}_k\left(P_S, P_{R_k}, \phi_{\sf opt}\right) = \left(\left|A\right| - \left|B\right|\right)^2.
\eeqn
\end{theorem}
The proof can be found in the technical report \cite{Techrep}.



\section{Power Control and Relay Section in the Non-coherent Scenario}
\label{PCRS_Configuration_NonCoh}

At the SU relay, we assume the self-interference is much higher than the noise, i.e., $\hat{\zeta} P_{R_k} >> \sigma_{R_k}^2$.
Therefore, we omit the term $\sigma_{R_k}^2$ in the object function.
Moreover $\log_2(1+x)$ is a strictly increase function in $x$, so we rewrite \textbf{Problem 1} as

\noindent
\textbf{Problem 2:}
\begin{equation}
\label{EQN_OPTRS_1}
\begin{array}{l}
 {\mathop {\max }\limits_{P_S, P_{R_k}}} \quad \mathcal{\bar C}_k(P_S, P_{R_k})  \\
 \mbox{s.t.}\,\,\,\, \mathcal{I}_k^{\sf non}\left(P_S, P_{R_k}\right) \leq \mathcal{\overline I}_{P}, 0 \leq P_S \leq P_S^{\sf max}, \\
 \quad \quad 0 \leq P_{R_k} \leq P_{R_k}^{\sf max},\\
 \end{array}\!\!
\end{equation}
where
\beqn
\mathcal{\bar C}_k(P_S, P_{R_k}) = \frac{\frac{P_{R_k} \left|h_{R_kD}\right|^2}{\sigma_D^2} \frac{P_S \left|h_{SR_k}\right|^2}{\hat{\zeta} P_{R_k}}}  {\mathcal{\bar A}},
\eeqn
$\mathcal{\bar A}$ is given as
\beqn\label{EQN_A_BAR}
\mathcal{\bar A} = 1+ \frac{P_{R_k} \left|h_{R_kD}\right|^2}{\sigma_D^2} + \frac{P_S \left|h_{SR_k}\right|^2}{\hat{\zeta} P_{R_k}}
\eeqn
and $\hat{\zeta}$ is calculated in (\ref{EQN_zeta_hat}).


\begin{lemma} \label{Lemma_noncoh1}
\textbf{Problem 2} is a nonconvex optimization problem for variables $\left(P_S, P_{R_k}\right)$.
\end{lemma}


\begin{lemma}
Given $P_S \in \left[0, P_S^{\sf max}\right]$, \textbf{Problem 2} is a convex optimization problem in terms of $P_{R_k}$. Similarly, given $P_{R_k} \in \left[0, P_{R_k}^{\sf max}\right]$, \textbf{Problem 2} is also a convex optimization problem in terms of $P_S$. \label{lem: nonco-convex}
\end{lemma}

The proofs for these lemmas are omitted and can be found in the technical report \cite{Techrep}.
Since \textbf{Problem 2} is non-convex, we exploit alternating-optimization problem (according to Lemma \ref{lem: nonco-convex}, the problem is convex when we fix one variable and optimize the other) to solve \textbf{Problem 2}, where each step is a convex optimization problem and can be solved using standard approaches \cite{Boyd04}. Finally, we determine the best relay by solving (\ref{EQN_OPT_RELAY_SELEC}).





We now consider the special case of ideal self-interference cancellation, i.e., $\hat{\zeta} = 0$.
We characterize the optimal solutions for \textbf{Problem 1} in the special case by the following lemma.

\begin{lemma} \label{Lemma_noncoh1_zeta0}
\textbf{Problem 1} is a convex optimization problem for variables $\left(P_S, P_{R_k}\right)$ when $\hat{\zeta} = 0$. 
\end{lemma}

The proof of Lemma ~\ref{Lemma_noncoh1_zeta0} is in our technical report \cite{Techrep}.
Based on Lemma ~\ref{Lemma_noncoh1_zeta0}, we can solve \textbf{Problem 1} when $\hat{\zeta} = 0$ by using fundamental methods \cite{Boyd04}.

\section{Power Control and Relay Selection in the Coherent Scenario}
\label{PCRS_Configuration_Coh}

Again, we assume that the self-interference is much higher than the noise at the selected relay, i.e., $\hat{\zeta} P_{R_k} >> \sigma_{R_k}^2$.
\textbf{Problem 1} can thus be reformulated as  

\noindent
\textbf{Problem 3:}
\begin{equation}
\label{EQN_OPTRS_2}
\begin{array}{l}
 {\mathop {\max }\limits_{P_S, P_{R_k}}} \quad \mathcal{\bar C}^{\sf coh}_k(P_S, P_{R_k})  \\
 \mbox{s.t.}\,\,\,\, \mathcal{I}_k^{\sf coh}\left(P_S, P_{R_k}\right) \leq \mathcal{\overline I}_{P}, 0 \leq P_S \leq P_S^{\sf max}, \\ 
 \quad \quad 0 \leq P_{R_k} \leq P_{R_k}^{\sf max},\\
 \end{array}\!\!
\end{equation}
where
\beqn
\mathcal{\bar C}^{\sf coh}_k(P_S, P_{R_k}) = \frac{\frac{P_{R_k} \left|h_{R_kD}\right|^2}{\sigma_D^2} \frac{P_S \left|h_{SR_k}\right|^2}{\hat{\zeta} P_{R_k}}}  {1+ \frac{P_{R_k} \left|h_{R_kD}\right|^2}{\sigma_D^2} + \frac{P_S \left|h_{SR_k}\right|^2}{\hat{\zeta} P_{R_k}}}
\eeqn
and $\hat{\zeta}$ is calculated in (\ref{EQN_zeta_hat}).

To solve \textbf{Problem 3}, the new variables are introduced as $p_S = \sqrt{P_S}$ and $p_{R_k} = \sqrt{P_{R_k}}$.
Hence \textbf{Problem 3} can be equivalently formulated as

\noindent
\textbf{Problem 4:}
\begin{equation}
\label{EQN_OPTRS_3}
\begin{array}{l}
 {\mathop {\max }\limits_{p_S, p_{R_k}}} \quad \mathcal{\breve{C}}^{\sf coh}_k(p_S, p_{R_k})  \\
 \mbox{s.t.}\,\,\,\, \mathcal{I}_k^{\sf coh}\left(P_S, P_{R_k}\right) \leq \mathcal{\overline I}_{P}, 0 \leq p_S \leq \sqrt{P_S^{\sf max}}, \\
 \quad \quad 0 \leq p_{R_k} \leq \sqrt{P_{R_k}^{\sf max}},\\
 \end{array}\!\!
\end{equation}
where the objective function is written as
\beqn
\mathcal{\breve{C}}^{\sf non}_k(p_S, p_{R_k}) = \frac{\frac{p_{R_k}^2 \left|h_{R_kD}\right|^2}{\sigma_D^2} \frac{p_S^2 \left|h_{SR_k}\right|^2}{\hat{\zeta} p_{R_k}^2}}  {1+ \frac{p_{R_k}^2 \left|h_{R_kD}\right|^2}{\sigma_D^2} + \frac{p_S^2 \left|h_{SR_k}\right|^2}{\hat{\zeta} p_{R_k}^2}}
\eeqn

We give a characterization of optimal solutions for \textbf{Problem 4} by the following lemmas.

\begin{lemma} \label{Lemma_coh1} \textbf{Problem 4} is not a convex optimization problem for variable $\left(p_S,p_{R_k}\right)$.\end{lemma}

\begin{lemma} Given $p_S \in \left[0, \sqrt{P_S^{\sf max}}\right]$, \textbf{Problem 4} is a convex optimization problem for variable $p_{R_k}$. Similarly, given $p_{R_k} \in \left[0, \sqrt{P_{R_k}^{\sf max}}\right]$, \textbf{Problem 4} is also a convex optimization problem for variable $p_S$. \label{lem: co-convex}\end{lemma}

The proofs of these lemmas can be found in our technical report \cite{Techrep}.
Based on Lemma \ref{lem: co-convex}, we again develop the alternating-optimization strategy to solve \textbf{Problem 4}, where each step is a convex optimization problem and can be solved using basic approaches \cite{Boyd04}. The relay selection is then determined by solving (\ref{EQN_OPT_RELAY_SELEC}).


We now investigate the special case of ideal self-interference cancellation, i.e., $\hat{\zeta} = 0$.
We then characterize the optimal solutions for \textbf{Problem 4} in the special case by the following lemma.

\begin{lemma} \label{Lemma_coh2_zeta0}
\textbf{Problem 4} is a convex optimization problem for variables $\left(P_S, P_{R_k}\right)$ when $\hat{\zeta} = 0$.
\end{lemma}

The proof of Lemma ~\ref{Lemma_coh2_zeta0} is in our technical report \cite{Techrep}.
According to Lemma ~\ref{Lemma_coh2_zeta0}, we can solve \textbf{Problem 4} in this special case by using standard approaches \cite{Boyd04}.

\vspace{10pt}
\section{Numerical Results}
\label{Results}

In the numerical evaluation, we chose the key parameters of the FDCRN as follows.
Each link is a Rayleigh fading channel with variance one (i.e., $\sigma_{SR_k}$ = $\sigma_{R_kD}$ = 1), except  $\sigma_{SD}$ = 0.1.
The noise power at every node is also set to be one.
The channel gains for the links of the SU relay-PU receiver and SU source-PU receiver are assumed to be Rayleigh-distributed with  variances $\left\{\sigma_{SP}, \sigma_{R_kP}\right\} \in \left[0.8, 1\right]$.
We also assume that the impact of imperfect channel estimation is included in only the parameter $\zeta$.

\begin{table*}
\centering
\caption{Achievable rate vs $\mathcal{\bar I}_P$ ($P_{\sf max} = 20 dB$)}
\label{table1}
\begin{tabular}{|c|c|c|c|c|c|c|c|}
\hline
\multicolumn{2}{|c|}{$\mathcal{\bar I}_P$ (dB)}
       & 0   & 2   & 4   & 6   & 8 & 10\tabularnewline
\hline
\hline
$\zeta = 0.001$,      & Optimal & 4.3646  &  5.1933  &  5.5533  &  5.6944  &  5.8162  &  5.9155 \tabularnewline
\cline{2-8}
Coherent  & Greedy  & 4.3513  &  5.1807  &  5.5496  &  5.6811  &  5.8131  &  5.8826  \tabularnewline
\cline{2-8}
scenario & $\Delta \mathcal{C} (\%)$ & 0.3047   & 0.2426   & 0.0666   & 0.2336   & 0.0533  &  0.5562  \tabularnewline
\hline
\hline
$\zeta = 0.001$,    & Optimal & 1.2390  &  1.6946  &  2.2118  &  2.7753  &  3.3718  &  3.9902  \tabularnewline
\cline{2-8}
Non-coherent  & Greedy  & 1.2309  &  1.6856  &  2.2018  &  2.7650  &  3.3610  &  3.9791  \tabularnewline
\cline{2-8}
scenario & $\Delta \mathcal{C} (\%)$ & 0.6538  &  0.5311  &  0.4521  &  0.3711  &  0.3203  &  0.2782  \tabularnewline
\hline
\end{tabular}
\end{table*}

We first demonstrate the efficacy of the proposed algorithms by comparing their achievable rate performances with those obtained by the optimal brute-force search algorithms.
Numerical results are presented for both coherent and non-coherent scenarios.
In Table~\ref{table1}, we consider the scenario with $\zeta$ = 0.001, 8 SU relays and $P_{\sf max} = 20$ dB.
We compare the achievable rate of the proposed and optimal algorithms for $\mathcal{\bar I}_P = \left\{0, 2, 4, 6, 8, 10\right\}$ dB.
These results confirm that our proposed algorithms achieve rate very close to that attained by the optimal solution (i.e., the errors are lower than 1\%).

\begin{figure}[!t]
\centering
\includegraphics[width=60mm]{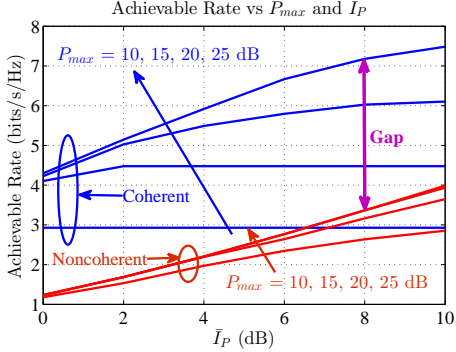}
\caption{Achievable rate versus the interference constraint $\mathcal{\bar I}_P$
 for $K = 8$, $\zeta = 0$, $P_{\sf max} = \left\{10, 15, 20, 25\right\}$ dB, and both coherent and non-coherent scenarios.}
\label{1_Rate_vs_I_bar_P_P_max_10152025_zeta_0}
\end{figure}

\begin{figure}[!t]
\centering
\includegraphics[width=60mm]{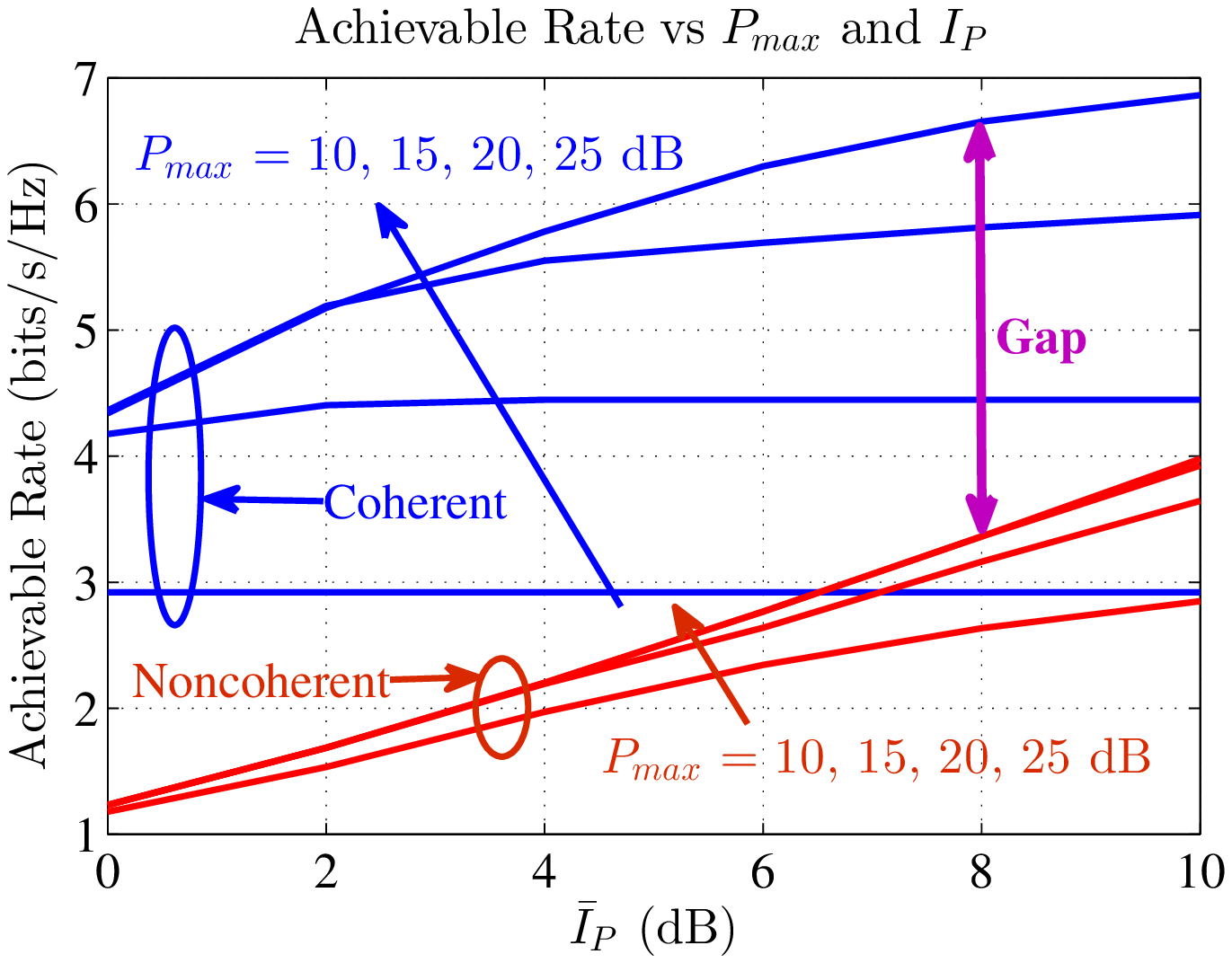}
\caption{Achievable rate versus the interference constraint $\mathcal{\bar I}_P$
 for $K = 8$, $\zeta = 0.001$, $P_{\sf max} = \left\{10, 15, 20, 25\right\}$ dB, and both coherent and non-coherent scenarios.}
\label{1_Rate_vs_I_bar_P_P_max_10152025_zeta_0001}
\end{figure}

\begin{figure}[!t]
\centering
\includegraphics[width=60mm]{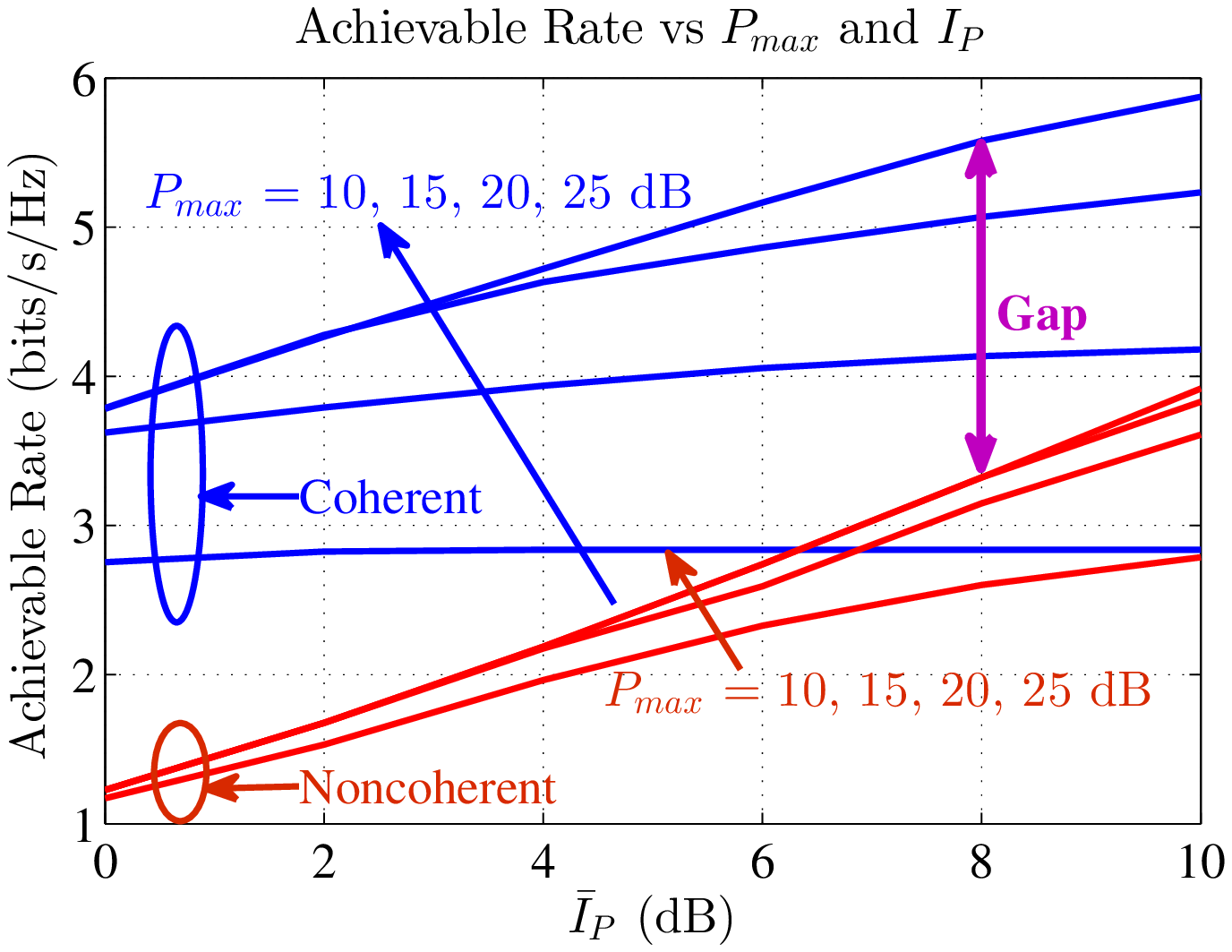}
\caption{Achievable rate versus the interference constraint $\mathcal{\bar I}_P$
 for $K = 8$, $\zeta = 0.01$, $P_{\sf max} = \left\{10, 15, 20, 25\right\}$ dB, and both coherent and non-coherent scenarios.}
\label{1_Rate_vs_I_bar_P_P_max_10152025_zeta_001}
\end{figure}

\begin{figure}[!t]
\centering
\includegraphics[width=60mm]{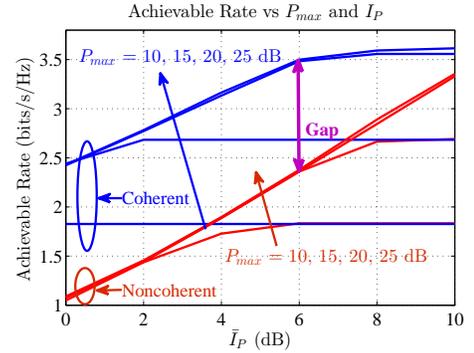}
\caption{Achievable rate versus the interference constraint $\mathcal{\bar I}_P$
 for $K = 8$, $\zeta = 0.4$, $P_{\sf max} = \left\{10, 15, 20, 25\right\}$ dB, and both coherent and non-coherent scenarios.}
\label{2_Rate_vs_I_bar_P_P_max_10152025_zeta_04}
\end{figure}

We then consider a FDCRN $8$ SU relays with $\zeta$ = 0, 0.001, 0.01, and 0.4, which represent ideal, high, medium and low Quality of Self-Interference Cancellation (QSIC), respectively. The tradeoffs between the achievable rate of the FDCRN and the interference constraint are shown in Figs.~\ref{1_Rate_vs_I_bar_P_P_max_10152025_zeta_0}, \ref{1_Rate_vs_I_bar_P_P_max_10152025_zeta_0001}, \ref{1_Rate_vs_I_bar_P_P_max_10152025_zeta_001} and \ref{2_Rate_vs_I_bar_P_P_max_10152025_zeta_04} under different values of $\zeta.$ In these numerical results,  we chose $P_S^{\sf max} = P_{R_k}^{\sf max} = P_{\sf max}$ for simplicity.


We have the following observations from these numerical results.
Firstly, the achievable rates of the coherent mechanism are always significantly higher than those of the  non-coherent mechanism.
This is because the phase is carefully regulated to reduce the interference at the PU receiver imposed by the SU transmissions, which allows higher transmit power both at the SU source and the SU relay.
Furthermore, the achievable rate decreases as expected when the QSIC increases due to the increase of self-interference at the FD relay.
\begin{figure}[!t]
\centering
\includegraphics[width=60mm]{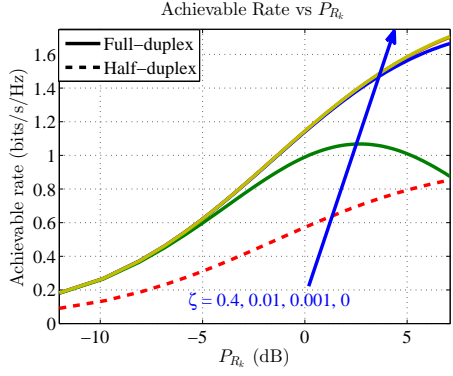}
\caption{Achievable rate versus the transmitted powers of SU relay $P_{R_k}$
for fixed $P_S = 5$ dB, $K = 10$, $\mathcal{\bar I}_P = 8$ dB, $P_{\sf max} = 25$ dB, and the non-coherent scenario.}
\label{Rate_vs_P_Rk_Non_P_S_5dB}
\end{figure}

\begin{figure}[!t]
\centering
\includegraphics[width=60mm]{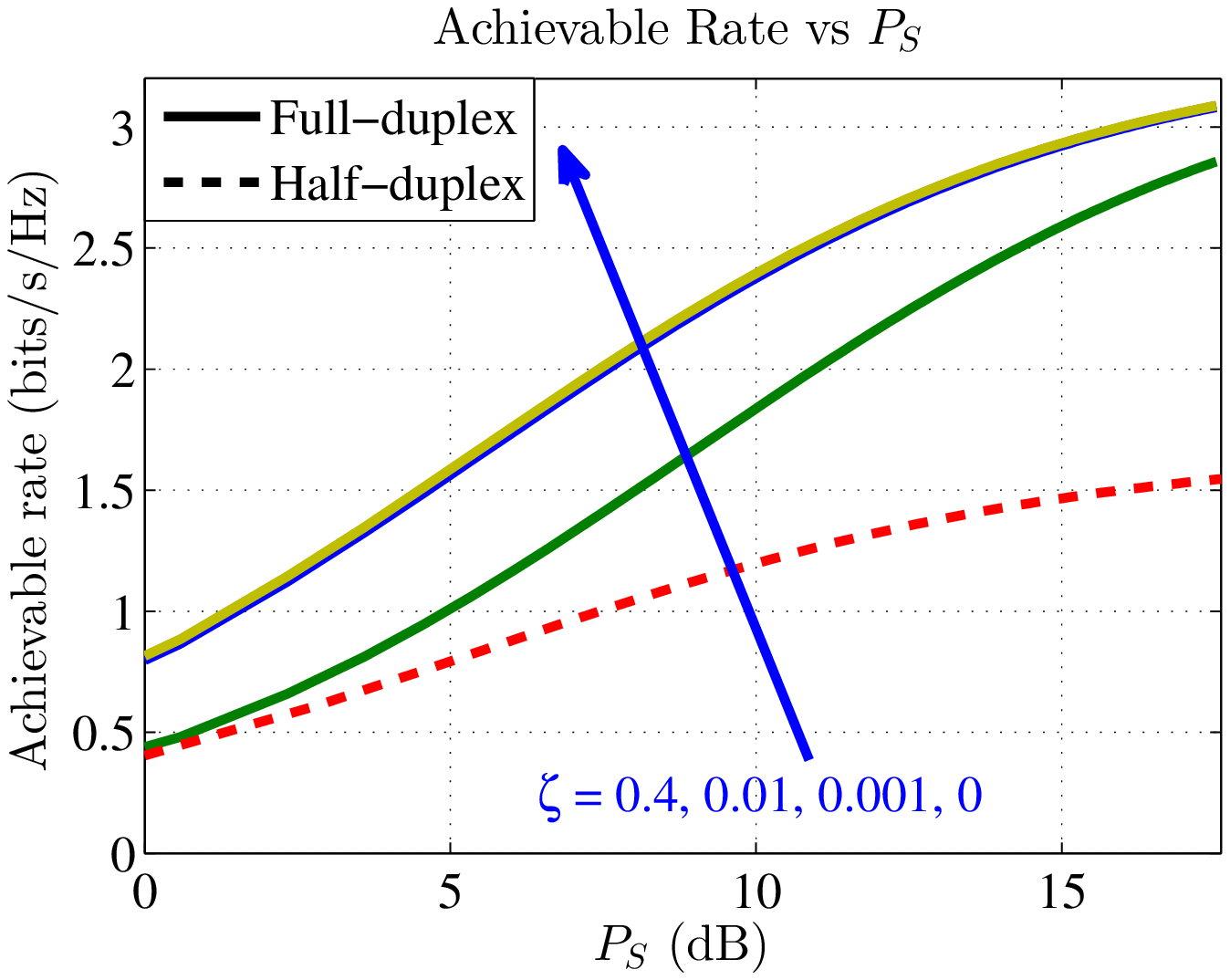}
\caption{Achievable rate versus the transmitted powers of SU source $P_S$
for fixed $P_{R_k} = 5$ dB, $K = 10$, $\mathcal{\bar I}_P = 8$ dB, $P_{\sf max} = 25$ dB, and the non-coherent scenario.}
\label{Rate_vs_P_S_Non_P_Rk_5dB}
\end{figure}

\begin{figure}[!t]
\centering
\includegraphics[width=60mm]{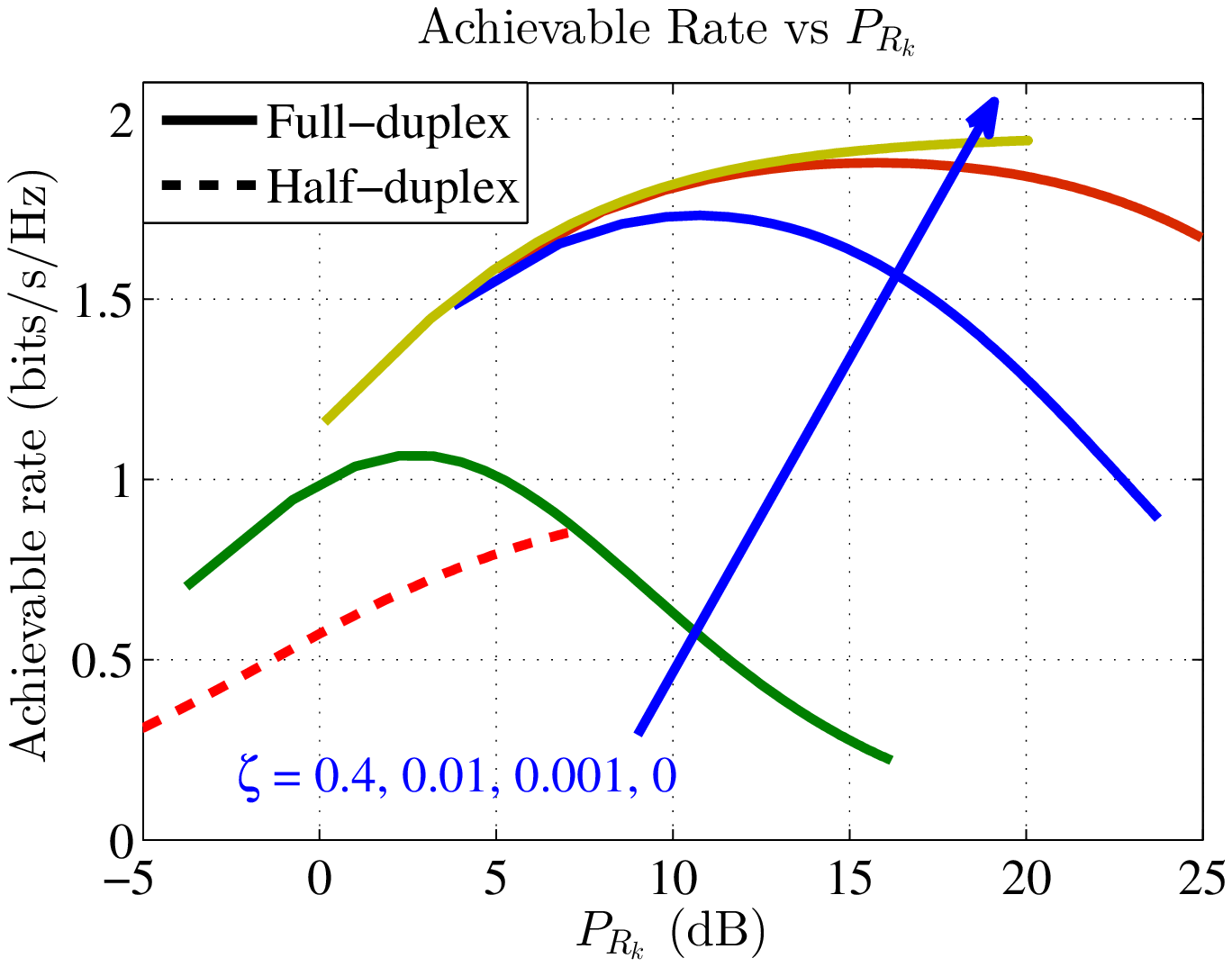}
\caption{Achievable rate versus the transmitted powers of SU relay $P_{R_k}$
for fixed $P_S = 5$ dB, $K = 10$, $\mathcal{\bar I}_P =8$ dB, $P_{\sf max} = 25$ dB, and the coherent scenario.}
\label{Rate_vs_P_Rk_Coh_P_S_5dB}
\end{figure}

\begin{figure}[!t]
\centering
\includegraphics[width=60mm]{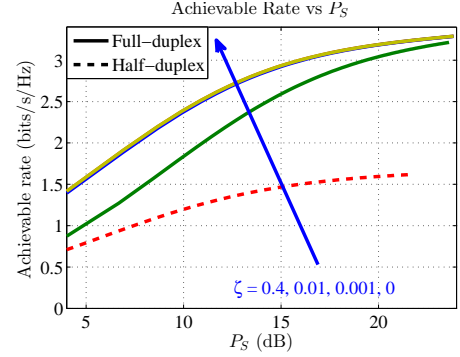}
\caption{Achievable rate versus the transmitted powers of SU source $P_S$
for fixed $P_{R_k} = 5$ dB, $K = 10$, $\mathcal{\bar I}_P =8$ dB, $P_{\sf max} = 25$ dB, and the coherent scenario.}
\label{Rate_vs_P_S_Coh_P_Rk_5dB}
\end{figure}

We now show the achievable rates of the FDCRN under different values of $P_{R_k}$ when fixing $P_S = 5$ dB in Fig.~\ref{Rate_vs_P_Rk_Non_P_S_5dB}.
The channel gains of the links of the SU relay-PU receiver and SU source-PU receiver were assumed to be Rayleigh-distributed with  variances $\left\{\sigma_{R_kP}, \sigma_{SP}\right\} \in \left[0.8, 1\right]$.
Fig.~\ref{Rate_vs_P_Rk_Non_P_S_5dB} evaluates the non-coherent scenario, $K = 10$ SU relays and $\mathcal{\bar I}_P =8$ dB.
All four cases (low, medium, high and ideal QSIC) have similar behaviors and achieve higher rate than the half-duplex case. For low QSIC (i.e., $\zeta = 0.4$),  the rate first increases then decreases as $P_{R_k}$ increases where the rate decrease is due to the strong self-interference.

Fig.~\ref{Rate_vs_P_S_Non_P_Rk_5dB} illustrates the achievable rates of the FDCRN against $P_S$ for fixed $P_{R_k} = 5$ dB in the non-coherent scenario. We considered $K = 10$ SU relays and $\mathcal{\bar I}_P = 8$ dB. The results from both Figs.~\ref{Rate_vs_P_Rk_Non_P_S_5dB} and \ref{Rate_vs_P_S_Non_P_Rk_5dB} confirm that the proposed power allocation for the FDCRN outperforms the HDCRN. 

We finally present the coherent scenario, $K = 10$ SU relays and $\mathcal{\bar I}_P = 8$ dB.
Fig.~\ref{Rate_vs_P_Rk_Coh_P_S_5dB} demonstrates the achievable rates of the FDCRN under different values of $P_{R_k}$ when fixing $P_S = 5$ dB; while Fig.~\ref{Rate_vs_P_S_Coh_P_Rk_5dB} demonstrates the achievable rates of the FDCRN under different values of $P_S$ when fixing $P_{R_k} = 5$ dB.
In the coherent scenario, we also have the same observations as those in the non-coherent scenario.

\section{Conclusion}
\label{conclusion}
This paper studied power control and relay selection in FDCRNs. We formulated the rate maximization problem, analyzed the achievable rate under the interference constraint, and proposed joint power control and relay selection algorithms based on alternative optimization.
The design and analysis have taken into account the self-interference of the FD transceiver, and included the both coherent and non-coherent scenarios. Numerical results have been presented to demonstrate the impacts
of the levels of self-interference and the significant gains of the coherent mechanism.

\section*{Acknowledgment}
This work was supported in part by the NSF under Grant CNS-1262329, ECCS-1547294, ECCS-1609202, and the U.S. Office of Naval Research (ONR) under Grant N00014-15-1-2169.

\bibliographystyle{IEEEtran}

\end{document}